\documentclass[preprint,aps,draft,floatfix,showpacs,preprintnumbers,amsmath,amssymb]{revtex4-1}

\usepackage{graphicx}% Include figure files
\usepackage{dcolumn}% Align table columns on decimal point
\usepackage{bm}% bold math
\usepackage{subfigure}
\usepackage{color}
\begin{document}
\preprint{RUP-12-13}

\title{
%Infinitely large energy extraction from extremal Reissner-Nordstr\"om black hole
%Extraction of arbitrarily high energy from extreme Reissner-Nordstr\"om black holes
%Extraction of Arbitrarily High Energy from Maximally Charged Black Holes
%via collisional Penrose process
Escape of superheavy and highly energetic particles produced by particle collisions near maximally charged black holes
}
\author{$^{1}$Hiroya Nemoto}\email{nemotoh@rikkyo.ac.jp}
\author{$^{1,2}$Umpei Miyamoto}\email{umpei@akita-pu.ac.jp}
\author{$^{1}$Tomohiro Harada}\email{harada@rikkyo.ac.jp}
\author{$^{1}$Takafumi Kokubu}
\date{\today}
\affiliation{$^{1}$Department of Physics, Rikkyo University, Toshima, Tokyo 171-8501, Japan\\
$^{2}$Research and Education Center for Comprehensive Science, Akita
Prefectural University, Akita 015-0055, Japan
}

\begin{abstract}
For particle 
collision near rapidly rotating Kerr black holes, 
the center-of-mass energy can be arbitrarily high 
if the angular momentum of either of the colliding particles is fine-tuned.
Recently, it 
has been shown that 
particles which are produced by 
such a particle collision and escape to infinity 
cannot be very massive nor very energetic. 
For electrically charged black holes there is a similar phenomenon, where 
the center-of-mass energy for the collision of charged particles 
near the horizon can be arbitrarily high. 
One might expect that there would exist a similar bound on the energy 
and mass of particles that are produced by such a particle collision 
and escape to infinity. 
In this paper, however, we see that this expectation is not the case. We 
explicitly show that superheavy and highly energetic charged particles produced by the 
collision near maximally charged black holes 
can escape to infinity at least within classical theory if the backreaction and self-force of the particle can be neglected.

\end{abstract}

\maketitle

%----------------------------------------%
%----------------------------------------%
\section{Introduction}
%----------------------------------------%
%----------------------------------------%

Ba\~nados, Silk, and West~\cite{Banados:2009pr} 
have shown that the center-of-mass (CM) energy of a particle collision
near the horizon of a maximally rotating black hole can be 
arbitrarily high if either of the two colliding particles has a
fine-tuned value of the angular momentum, which we call 
the Ba\~nados-Silk-West (BSW) collision.
Based on this demonstration, they have suggested 
that a maximally rotating Kerr black 
hole acts as a particle accelerator, 
so that a maximally rotating Kerr black hole is expected to create quite 
a massive or highly energetic particle in principle. 
If such an exotic particle is to be observed at infinity, it 
must escape to infinity. Jacobson and 
Sotiriou~\cite{Jacobson:2009zg} have suggested that 
such a massive or energetic particle cannot escape to infinity.

For this problem, the energy extraction mechanism from a black hole 
might be relevant. Penrose considered the disintegration of an incident 
particle in the ergoregion and showed that 
energy can be extracted from a rotating black hole 
{with the production of} a negative-energy particle in the 
ergoregion~\cite{Penrose:1969pc}. This process is called 
the Penrose process. 
The energy extraction from a rapidly rotating black hole is expected to 
be a possible engine of active galactic nuclei. 
On the other hand, the net energy-extraction efficiency of the Penrose process 
is bounded by the upper limit $ \simeq 
20.7$\%~\cite{Bardeen:1972fi,wald,Chandrasekhar}. 
Piran, Shaham, and
Katz~\cite{Piran_etal_1975,Piran:1977dm,Piran_Shaham_1977_grb} proposed
a collisional Penrose process. They first expected that it is much more
efficient than the original Penrose process, whereas they subsequently 
found that its efficiency is as modest as the original one. 
Very recently, Bejger, Piran, Abramowicz, and H\r{a}kanson~\cite{Bejger:2012yb} have 
numerically shown that the energy of a photon produced by the pair 
annihilation of the BSW type cannot be high, but instead rather modest. 
Harada, Nemoto, and Miyamoto 
have analytically given upper bounds on both the energy of the 
emitted particle and the energy-extraction efficiency for more general physical 
reactions of the BSW type~\cite{Harada:2012ap}. 
Both these studies have revealed that 
the collisional Penrose process through the BSW collision is a valid 
energy-extraction mechanism, while the emitted particle escaping 
to infinity cannot be very energetic nor very massive.

As for an electrically charged black hole, 
Denardo, Hively, and Ruffini showed that there exists a region of the
spacetime where the energy of a charged particle can be negative and
hence the energy can be extracted~\cite{Denardo:RN,Denardo:KN}. So the
Penrose process can occur around a charged black hole.
In fact, the energy extraction from a black hole with an electromagnetic field 
can be more efficient than that from a neutral rotating black hole and its 
efficiency is not bounded~\cite{Wagh.mag:1985,Dhurandhar.a:1984,Dhurandhar.b:1984,Bhat.KN:1985,Parthasarathy:1986}. 
Recently, Zaslavskii~\cite{Zaslavskii:2010aw} has shown 
that there is an electromagnetic counterpart 
of the BSW collision.
In the case of a static maximally charged black hole, 
a collision of arbitrarily high CM energy is possible 
if either of the two colliding particles has a fine-tuned value of 
the charge. 
One can expect that a more efficient energy extraction 
through the 
collision of this type near a Reissner-Nordstr\"om black hole may be
possible than that through the BSW collision near a Kerr black hole. 
We show that this is the case.  
Unlike in the Kerr case, we explicitly demonstrate 
that the mass and energy of the product particle 
which escapes to infinity can be arbitrarily large.
Note that, in the same context, Zaslavskii~\cite{Zaslavskii:2012ax} has 
firstly shown that there exists no upper bound on the energy extraction
from the collision of this type and noticed the fundamental 
difference in this regard between the Kerr case and the
Reissner-Nordstr\"om case.
His approach and our present approach are totally consistent with and 
complementary to each other.
This phenomenon, in principle, opens a possibility that a distant observer
might observe an exotic superheavy particle produced by the particle 
collision of extremely high CM energy near a maximally 
charged black hole.  
We use the geometrized units, in which $ c = G = 4\pi \epsilon_0=1$.

%\section{Charged particle around a Reissner-Nordstr\"om black hol}
\section{Charged particle around a Reissner-Nordstr\"om black hole}

We briefly review the motion of a test charged particle in a Reissner-Nordstr\"om black hole spacetime. The line element in the Reissner-Nordstr\"om spacetime is given by 
		\begin{equation}\label{metric}
		ds^2 = -f(r) dt^2  +  \frac{1}{f(r)} dr^2  +  r^2 d\theta^2  +  \sin ^2\theta d\phi ^2
		,
		\end{equation}
where
		\begin{equation}
		f(r)  =  1  -  \frac{2M}{r}  +  \frac{Q^2}{r^2}
		,
		\end{equation}
and $M$ and $Q$ ($|Q| \leq M$) are the mass and electric charge of the 
black hole, respectively. 
We assume $Q\ge 0$ hereafter without loss of generality. An event horizon is
located at $r = r_h \equiv M + \sqrt{M^2 - Q^2}$, where $f(r)$
vanishes. 
If $Q=M$, $r_h=M$ and the black hole is said to be extremal.

%-----------------------------------------------%
%%% the Lagrangian of a charged particle
%-----------------------------------------------%
The Lagrangian of a test charged particle in an electromagnetic field is given by 
		\begin{equation}
		\mathcal L = \frac{1}{2} g_{\mu\nu} \frac{d x^\mu}{d \lambda} \frac{d x^\nu}{d \lambda} + q A_\mu \frac{d x^\mu}{d \lambda}
		,
		\end{equation}
where $q$ is the electric charge of the particle, $\lambda$ is the
parameter of the particle's 
world line, and $A_\mu = -Q/r(dt)_\mu$ is a vector potential. 
The parameter $\lambda $ of the particle with mass $m$ is related to the 
proper time $\tau $ by $\tau =m\lambda $. The local four-momentum 
$p^\mu$ is given by $p^\mu=dx^\mu/d\lambda $, where $p^\mu$ is 
normalized as $p^\mu p_\mu=-m^2$. If we put $m=0$, we obtain the 
motion of a massless particle. Since the Reissner-Nordstr\"om 
spacetime is spherically symmetric, we can assume the motion of a 
particle is restricted on the equatorial plane 
($\theta=\pi/2$). Since the metric~(\ref{metric}) does not depend on 
$t$ or $\phi$, 
we obtain from the Euler-Lagrange equation 
\begin{equation}
\frac{dt}{d \lambda} = \frac{P(r)}{f(r)}
\end{equation}
and
\begin{equation}
\frac{d\phi}{d \lambda} = \frac{L}{r^2}
,
\end{equation}
where
\begin{equation}\label{6}
P(r) \equiv E - q\frac{Q}{r}
,
\end{equation}
and $E$ and $L$ denote the conserved energy and angular momentum
of the particle, respectively. The equation of the radial motion 
is written in the following form:
%-----------------------------------------------%
%%% effective potential
%-----------------------------------------------%
		\begin{equation}
		\left( \frac{dr}{d\lambda} \right)^2 + V(r)  =  0,
		\end{equation}
where the effective potential $V(r)$ is given by 
		\begin{equation}\label{8}
		V(r)  =  -\left( E - \frac{qQ}{r} \right)^2 +  f(r) \left( m^2 + \frac{L^2}{r^2} \right)
		.
		\end{equation}
In the region where the motion of the particle is allowed, $V(r)$ must
be nonpositive. In the limit $r\to \infty $, $V(r)$ becomes 
		\begin{equation}\notag
		\lim_{r\to \infty} V(r)  =  -E^2 + m^2,
		\end{equation}
implying that the particle can be at infinity if $E\geq m$.

%-----------------------------------------------%
%C forward-in-time condition
%-----------------------------------------------%
Since we are considering a causal world line, we need to impose
$dt/d\lambda \geq 0$ along it. This is called a
forward-in-time condition. Since we are interested in the outside of 
the horizon, the
following inequality must hold in the region in which the particle 
is allowed to exist:
\begin{equation}\label{fit}
E - \frac{qQ}{r} \geq 0.
\end{equation}
In particular, to reach the horizon from the outside of the horizon, the particle must satisfy 
\begin{equation}
\label{cri}
E \geq \frac{qQ}{r_H}
.
\end{equation}
We call a particle which satisfies the equality in Eq.~(\ref{cri}) a
critical particle and the equality a critical condition.

%-----------------------------------------------%
%%% zero-AM 運動
%-----------------------------------------------%
In the following, we consider particles with vanishing angular momentum,
i.e., $L=0$, in the maximally charged black hole spacetime. 
There are two turning points, where $V(r)$ vanishes, given by 
\begin{equation}\label{turn}
r = r_\pm \equiv M\left( 1 + \frac{q- E}{E \mp m} \right).
\end{equation}
If the particle is unbounded or $E>m$, $r_+$ is an outer turning point. 
If an unbounded particle satisfies $q> E$, $M<r_{-}<r_+$ holds and 
the region between $r_{-}$ and $r_{+}$ is prohibited.

%----------------------------------------%
%----------------------------------------%
\section{Collision of arbitrarily high CM energy}
%----------------------------------------%
%----------------------------------------%
We consider two charged particles falling radially into the maximally 
charged black hole. 
The CM energy $E_{\rm cm}$ 
of the two colliding particles in the limit where the collision point approaches the horizon is given by
\begin{equation}\label{ecmh}
\lim_{r\to r_H} E_{\rm cm}^2
=
m_1^2 + m_2^2 +
\left[
\frac{E_2 - q_2}{E_1 - q_1} m_1^2
+ \frac{E_1 - q_1}{E_2 - q_2} m_2^2
\right]
.
\end{equation}
The constants $E_i$, $m_i$, and $q_i$ ($i=1,2$) denote the
conserved energy, mass, and electric charge of particle $i$,
respectively. If either of the two particles 
satisfies $E_i=q_i$ or the critical condition, the right-hand side of
Eq. (\ref{ecmh}) diverges. 
If the critical particle satisfies $E_i>m_i$, the potential does not
prevent the critical particle from approaching the horizon. If particle
1 is critical
and particle 2 is subcritical, i.e., $q_{1}=E_{1}$ and $q_{2}<E_{2}$, the
CM energy of the two particles colliding at $r=r_c\equiv M(1+\epsilon)$
($\epsilon \ll 1$) is given by 
\begin{equation}\label{lecm}
E_{\rm cm}^2 \approx \frac{2A (E_2 - q_2)}{\epsilon}
,
\end{equation}
where $A\equiv E_1 - \sqrt{E_1^2 - m_1^2}$. We can see from
Eq.~(\ref{lecm}) that the CM energy can grow without bound in the limit where the collision point approaches the horizon.

Now, we consider the reaction of particles 1 and 2 into particles 3 and
4, all of which are assumed to move radially, i.e., $L_i = 0$ ($i = 1, 2, 3,
4$). We assume that 
particle 4 is moving inwardly at the collision point, %on production, 
i.e., $p_4^r <0$.

The conservation of charge and four-momentum before and after the
collision at the collision point $r=r_c\equiv M(1+\epsilon)$ yields
		\begin{equation}\label{q-cons}
		q_1 + q_2 = q_3 + q_4
		\end{equation}
and 
		\begin{equation}\label{p-cons}
		p_1^\mu  +  p_2^\mu  =  p_3^\mu  +  p_4^\mu,
		\end{equation}
respectively.
Since all particles move radially, the $\theta$ and $\phi$ components of
Eq.~(\ref{p-cons}) vanish. Using Eq.~(\ref{q-cons}), we obtain the
energy conservation from the $t$ component of Eq.~(\ref{p-cons}). Since
the product particles after the collision are timelike or null,
particles 3 and 4 must satisfy $m_i^2\geq 0$ ($i=3,4$) and
Eq.~(\ref{fit}) in the region where they are allowed to exist. For
particle 3, the outer turning point must be located inside the 
collision point so that it can escape to infinity. 

%----------------------------------------%
%----------------------------------------%
\section{Escape of superheavy and highly energetic particles}
%----------------------------------------%
%----------------------------------------%
Here, we explicitly show that particle 3 can be very massive and escape to
infinity, simultaneously. For this purpose,  
we assume that particle 3 is moving inwardly or $p_3^r<0$ is satisfied 
at the collision point %on production 
and satisfies the following equations:
\begin{equation}\label{ncc}
q_3 = E_3 (1 + \delta \epsilon)
\end{equation}
and
\begin{equation}\label{asu1}
E_3(1-\delta) = \frac{1}{2} (A + \frac{m_3^2}{A})
,
\end{equation}
where $\delta$ is a constant between 0 and 1. $E_3 > m_3$ is 
automatically satisfied, which is necessary 
for particle 3 to escape to infinity. 
In the following, we check that particle 3 can be very massive 
and energetic and really escape to infinity, simultaneously.

First, we check the location of the outer turning point for particle 3.
Using Eqs.~(\ref{turn}) and (\ref{ncc}), we find that the outer turning point
for particle 3 is given by
\begin{equation}\label{oturn3}
r_{+,3} = M \left( 1 + \frac{E_3}{E_3 - m_3} \delta \epsilon \right)
.
\end{equation}
Since we assume $p_3^r<0$, particle 3 must be bounced by the potential
barrier to escape to infinity. The outer turning point $r_{+,3}$ is
located outside the horizon. 
From Eq.~(\ref{oturn3}), the requirement $r_c \geq
r_{+,3}$ is equivalent to 
\begin{equation}\label{turn3}
E_3 (1-\delta) \geq m_3.
\end{equation}
We find that the above inequality is automatically satisfied 
from Eq.~(\ref{asu1}).

Then, we check the forward-in-time condition for particle 3. Using
Eqs.~(\ref{ncc}) and (\ref{oturn3}), we can see that Eq.~(\ref{fit}) is automatically satisfied 
at $r=r_{+,3}$, 
and it is also automatically satisfied in the outer region.

Then, we check that all of $m_{2}^{2}\ge 0$, $m_{3}^{2}\ge 0$, and $m_{4}^{2}\ge
0$ can be satisfied.
The reaction must satisfy the local momentum conservation. 
The $r$ component of the four-momentum of particle
$i$ is given by $|p_i^r|=\sqrt{-V_i}$. Using Eqs.~(\ref{ncc}) and
(\ref{asu1}), we can solve the $r$ component of Eq.~(\ref{p-cons}) for
$m_3^2$. The result is the following:
\begin{equation}\label{m3ex}
m_3^2
=
A \frac{m_2^2 - m_4^2}{P_2 - \sqrt{-V_2}} \left( 1 - \frac{M}{r} \right)
+ m_1^2
,
\end{equation}
where $P_2$ and $V_2$ are given by Eqs.~(\ref{6}) and (\ref{8}) of particle 2,
respectively. Therefore, if $m_2^2>m_4^2\geq 0$, the right-hand side of
Eq.~(\ref{m3ex}) is positive. Thus, both $m_3^2$ and $m_4^2$ can be
positive. Especially, from Eq.~(\ref{m3ex}), $m_3^2$ for the
near-horizon 
collision ($\epsilon \ll 1$) is given by 
\begin{equation}\label{m3app}
m_3^2
\approx
\frac{2(E_2 - q_2)A }{\epsilon}
\left( 1 - \frac{m_4^2}{m_2^2} \right)
\approx
E_{\rm cm}^2 \left( 1 - \frac{m_4^2}{m_2^2} \right)
,
\end{equation}
where Eq.~(\ref{lecm}) has been used.
Therefore, if $m_2 > m_4$ is satisfied, $m_3^2$ is positive. The
above equation also implies that $m_{3}$ is typically of the order of 
the CM energy.

Finally, we check the forward-in-time condition for particle 4. Since
particle 4 moves inside the collision point, we study the left-hand side
of Eq.~(\ref{fit}) for particle 4 between the collision point and the
horizon. From the energy and charge conservation, $E_4$ and $q_4$ are
written in terms of $E_1$, $E_2$, $E_3$, $q_1$, $q_2$, and $q_3$. 
To estimate the left-hand side of Eq.~(\ref{fit}) for particle 4 
at the horizon, using the energy and charge conservation and
Eq.~(\ref{ncc}), we find
		\begin{equation}\label{fit4h}
		E_4 - q_4  =  E_2 - q_2 + E_3 \delta \epsilon
		.
		\end{equation}
Since $E_3>0$, $E_2>q_2$, and $\delta >0$, the right-hand side of
Eq.~(\ref{fit4h}) is positive. To estimate the left-hand side of
Eq.~(\ref{fit}) of particle 4 at the collision point, using the energy
and charge conservation, Eqs.~(\ref{ncc}), (\ref{asu1}), and
(\ref{m3app}),
we find
		\begin{equation}\label{fit4c}
		E_4 - q_4 \frac{1}{1+\epsilon}
		=
		\frac{1}{2} \left[ P_2 - \sqrt{P_2^2 - m_2^2 f(r)} + \frac{m_4^2}{m_2^2} \left( P_2 + \sqrt{P_2^2 - m_2^2 f(r)} \right) \right]
		\biggr|_{r=M(1+\epsilon)}
		.
		\end{equation}
Thus, if $m_4^2\geq 0$, the right-hand side of Eq.~(\ref{fit4c}) is
positive. Since the left-hand side of Eq.~(\ref{fit}) is a monotonic
function, particle 4 satisfies the forward-in-time condition in the
relevant region.

In the above, we have shown that the assumptions~(\ref{ncc}) and
 (\ref{asu1}) are consistent with the forward-in-time condition, $r_c >
 r_{+,3}$, the momentum conservation, $m_3^2\geq 0$,
 $m_4^2\geq 0$, and $E_3> m_3$. Now, we examine how energetic and
 massive particle 3 can be if $\epsilon \ll 1$. For simplicity, we 
assume $m_1=E_2=m_2=m$ and $q_2=0$. Since particle 1 is assumed to be
 critical, $E_1=q_1$ must be satisfied. For simplicity, we assume that $E_1$ is
 slightly larger than the rest mass energy $m_1$, while an exactly
 marginally bound critical particle, i.e., $E_{1}=q_{1}=m_{1}$, must be
 at rest. Since $m_4$ must be smaller than $m_2$, let us choose
 $m_4=m/\sqrt{2}$. If particles 1 and 2 collide at
 $r=M(1+\epsilon)$ ($\epsilon=10^{-4}$), we can find from Eq.~(\ref{m3app}) that the mass of
 particle 3 is given by $m_3\simeq 100m$. Since the constant $\delta $ can
 take any value between zero and unity, let us choose
 $\delta=1/2$. Using Eqs.~(\ref{ncc}) and (\ref{asu1}), we can find $E_3
 \simeq q_3\simeq 10^{4}m$. In general, for $\epsilon\ll 1$, 
we find that $m_{3}\sim E_{\rm cm}\sim
 m/\sqrt{\epsilon}$ and $E_{3}\sim q_{3}\sim m/\epsilon$
up to numerical factors of order unity.

%----------------------------------------%
%----------------------------------------%
\section{conclusion and discussion}
%----------------------------------------%
%----------------------------------------%

We have shown that a superheavy and highly energetic
particle, which can be produced by the collision near an extremal
Reissner-Nordstr\"om black hole,
can escape to infinity and be in principle observed by a distant observer. 
In other words, kinematics does not forbid
such a massive particle from escaping to infinity 
in contrast to the Kerr case, where kinematics forbids such a superheavy 
particle from escaping to infinity. %~\cite{Harada:2012ap}. 
In this study, we have neglected gravitational and electromagnetic radiation
and backreactions on the metric and the electromagnetic field. Thus, one 
may expect that there is an upper bound on the energy and 
mass of escaping product particles in realistic situations due to these
effects.

Finally, we should comment on the present mechanism in the context of 
particle physics in the real world.
It is well known that electric charge is quantized by an elementary 
charge $e\simeq 1.6\times 10^{-19}$C. The energy of the critical
particle in this paper is therefore discretized by the mass corresponding 
to $e$, which is given by $\sqrt{\alpha}E_{\rm Pl}$, where $\alpha\simeq
1/137$ is the fine-structure constant and $E_{\rm Pl}\simeq 10^{19}$GeV is 
the Planck energy.
This clearly makes it problematic to interpret the present mechanism as a 
factory of superheavy elementary particles in a usual sense.
To circumvent this problem, one might introduce a macroscopic object 
which satisfies a critical condition.
For example, an object as massive as $10^{-6}$ g can be critical 
if it is charged with only one elementary charge.  

%----------------------------------------%
%----------------------------------------%
\acknowledgments{T.H. would like to thank T.~Igata and M.~Kimura for
fruitful discussion.
U.M.\ was supported by the Research Center for
Measurement in Advanced Science in Rikkyo University. 
U.M. and T.H. were supported
by Grant-in-Aid for Scientific Research from the Ministry of Education,
Culture, Sports, Science, and Technology of Japan [Grants 
Nos. 25800157 and 21740190, respectively]. }
%----------------------------------------%
%----------------------------------------%

%----------------------------------------%
%----------------------------------------%

%----------------------------------------%
%----------------------------------------%

\end{document}